\documentclass[preprint,a4paper,10pt,3p, twocolumn]{elsarticle}

\usepackage{hyphenat}
\usepackage{amssymb}

\usepackage{lineno}
\usepackage{float}

\newdefinition{rmk}{Remark}

\usepackage{amsmath,amssymb,amsfonts,amsthm}
\usepackage{bbm}

\biboptions{sort&compress}
\usepackage{physics}

\usepackage{xcolor}
\usepackage{hyperref}
\usepackage{psfrag}

\usepackage{algorithm}
\usepackage[end]{algpseudocode}

\usepackage{wrapfig}
\usepackage{subcaption}

\usepackage{pgfplots}
\usepackage{pgfplotstable}
\pgfplotsset{
tick label style={font=\footnotesize},
label style={font=\small},
legend style={font=\small},
axis lines=box,
unbounded coords=jump,
scale only axis,
legend style={draw=black,fill=white,legend cell align=left},
every axis plot/.append style = {line width = 0.5pt},
ylabel near ticks,
xlabel near ticks,
compat = 1.3
}

\newcommand{\beginsupplement}{%
        \setcounter{table}{0}
        \renewcommand{\thetable}{S\arabic{table}}%
        \setcounter{figure}{0}
        \renewcommand{\thefigure}{S\arabic{figure}}%
        \setcounter{section}{0}
        \renewcommand{\thesection}{S\arabic{section}}
     }

\journal{Progress in Additive Manufacturing}

\begin{document}

\begin{frontmatter}





\title{A rapid experimental workflow for studying melt track scaling in laser powder bed fusion using high-precision metal template substrates}

\author[1]{Reimar~Weissbach}

\author[1]{Ryan~W.~Penny}


\author[1]{A.~John~Hart\corref{cor}}
\ead{ajhart@mit.edu}

\cortext[cor]{corresponding author}

\address[1]{Department of Mechanical Engineering, Massachusetts Institute of Technology, 77 Massachusetts Avenue, Cambridge, 02139, MA, USA}


\onecolumn
\begin{abstract}

Development and qualification of process parameters in laser powder bed fusion (LPBF) commonly involves many variables. At the outset of development, whether transferring known parameters to a new machine, or exploring a new material, single-track and single-layer experiments are a convenient means of down-selecting key variables and exploring parameter scaling behavior.  We present an experimental workflow for single-layer LPBF experiments using etched high-precision metal template substrates, overcoming challenges with precision single-layer alignment in LPBF systems and enabling efficient processing and cross-sectional analysis. Templates are fabricated using chemical etching and machining, and are characterized using optical profilometry and X-ray transmission imaging of powder layers. Using the templates a single-track parameter study is performed in SS316 including three powder layer thicknesses, and spanning common laser melting modes (lack-of-fusion, conduction, and keyhole mode). Analysis of melt track geometries using automated image processing allows a scaling law to be applied to define the process window, quantifying the amount of material added with increasing powder layer thickness. Single-track results are verified with raster scanning experiments, showing the potential to transfer single-track results to full LPBF builds.

\end{abstract}


\twocolumn
\end{frontmatter}



\section{Introduction} 
\label{sec:intro}

Laser powder bed fusion (LPBF) allows fabrication of near net shape parts with complex geometries without the need for specialized tooling. It is used in commercial applications such as jet engine components, orthopedic implants, and industrial tooling, among others~\cite{arabnejad2017fully,3DPrint2022}. In LPBF, key process parameters include laser power, laser spot size, scan speed, powder layer thickness, and hatch spacing. Given that the parameter space has a high dimensionality, one efficient approach is single-track and single-layer experiments, which can identify parameters that are down-selected for subsequent builds of three-dimensional specimens.


Single-track and single-layer LPBF parameter studies generally seek better understanding of: (i) powder flowability and powder layer formation during the spreading process~\cite{oropeza2022mechanized, chen2022high, Penny2021, PENNY2024Blade, PENNY2024Roller, weissbach2024exploration} where different kinematic parameters of the spreading tool (e.g., speed, rotation) can be applied; (ii) melting process parameters~\cite{nayak2020effect, nayak2021effect, martucci2021automatic, makoana2018characterization, matthews2017denudation, simson2024experimental, li2012balling, chen2022high, kouprianoff2017line, yuan2022understanding, zhang2020experimental, yadroitsev2010selective, nie2023effect, hanemann2020dimensionless, kosiba2025fabrication} to understand melt track scaling, the underlying physics (such as powder denudation) and processing regimes (i.e., balling, lack-of-fusion, conduction, keyholing); (iii) processing of new alloys and variants thereof~\cite{ghasri2023single}; (iv) microstructure formation as related to material composition and process parameters~\cite{mohammadpour2022microstructure, nie2023effect}; as well as (v) new modes of processing such as formation of porous materials~\cite{jafari2020porous}.  Single layer experiments also provide clear insight to the intersection of substrate and powder properties, by enabling localized investigation of the interface which is necessary for integrating LPBF onto pre-manufactured components or combining process technologies (e.g., LPBF onto machined surfaces).


However, it is difficult to achieve accurate single-layer data within conventional LPBF setups, because of the required flatness and parallelism of the build plate to the spreading tool. A rule of thumb of 10 $\mu$m flatness and parallelism, or 10-20\% of a typical build layer, is below the manufacturing tolerances typically achieved by grinding (e.g., blade and build plate) and therefore in a build the first several layers are used to allow the layers to self-planarize.  Moreover, to improve repeatability of experimental results, it is critical to understand and document the substrate conditions, such as surface finish (e.g., sandblasted, machined, etc.), flatness, or parallelism of the substrate to the spreading tool (many smaller machines do not have a leveling mechanism), as well as the spreading methodology. Single-layer data is often gathered from the top layer of multi-layer builds in order to avoid uncertainties stemming from the substrate, which requires more powder, and introduces other uncertainties as to geometry and surface effects. 

This paper presents a methodology for performing melt track experiments using high-precision metal templates, and applies this methodology to a melt track parameter study. Templates are fabricated using chemical etching and machining, and are compared using optical profilometry as well as X-ray transmission imaging of manually spread powder layers. An automated analysis tool is used to process cross-sectional images and reveal the scaling of melt track dimensions and geometry across a wide range of laser parameters spanning three layer thicknesses representative of LPBF implementations.

\section{Materials and Methods}
\subsection{Fabrication of single-layer templates}

\begin{figure*}
 \begin{center}
   \includegraphics[scale=1, keepaspectratio=true, width=\textwidth]{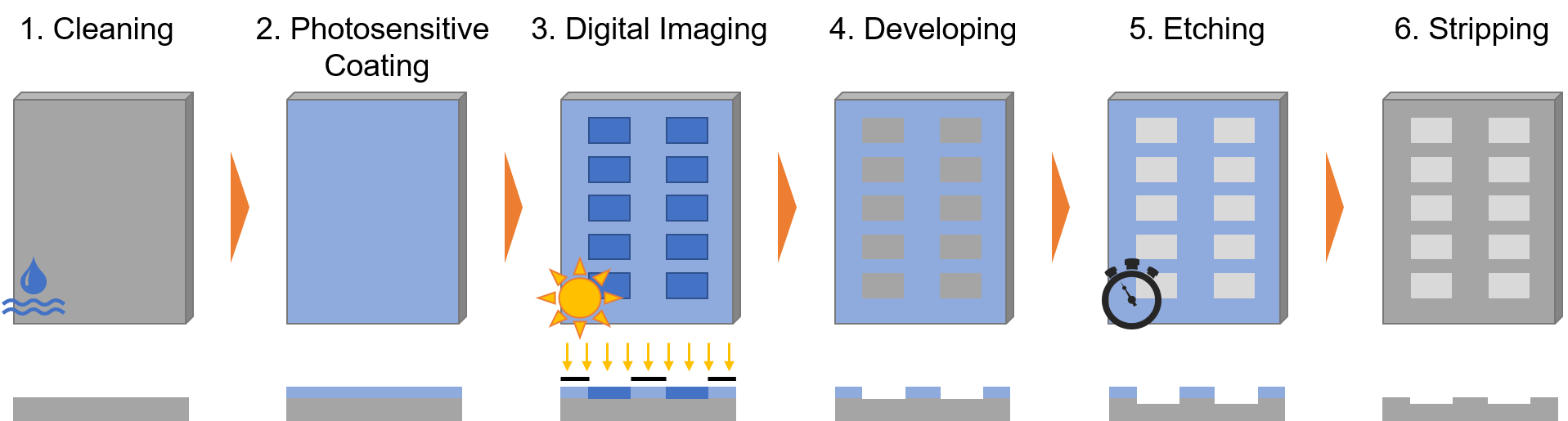}
 \end{center}
 \caption{Schematic of the etching process}
 \label{fig:etching process}
\end{figure*}

Shallow templates are created in a thin metal plate that can be mounted to the build plate in a LPBF machine. Template preparation as described in the following aims to achieve three primary objectives: (i) limited powder usage; (ii) uniform and repeatable powder layer and melt track quality; and (iii) efficient and convenient melt track analysis.  As such, two methods for manufacturing templates, namely chemical etching and machining of a SS316 substrate (1/8 inch thickness, McMaster-Carr part number 9090K22), are presented here.  

With chemical etching, template cavities with precise and uniform depth, and low surface roughness can be achieved from a suitably chosen substrate.  This approach was first demonstrated by Penny and Hart in~\cite{Penny2021}, wherein potassium hydroxide was used to etch templates in silicon wafers to sub-micron depth accuracy, which were used for spreading powder layers for analysis by X-ray microscopy. To bring a suitable level of precision to melting experiments, where the powder and substrate materials must be similar, chemical etching is adapted to 316 stainless steel substrates.  Crude templates can be produced using hand-cut kapton tape (McMaster-Carr P/N 2271K72) to mask $0.12$~in. thick 316 SS specimens (cut from McMaster-Carr P/N 88885K76 or equivalent) followed by etching in a glass beaker with an as-supplied 40~weight percent aqueous ferric chloride solution (MG Chemicals).  Etch rate is reactant-limited and therefore strongly dependent on the degree of solution agitation; etch rates of $1$~\textmu m/min are typical using a stir plate set to low speed and at room temperature, and etch depth is simply proportional to time.   

To achieve the desired precision, etched templates studied here are supplied commercially (United Western Enterprises), using a photolithographic process (see Figure~\ref{fig:etching process}.  After first cleaning the metal substrate, a photoresist coating is applied. The coated substrate coating is selectively exposed to UV light to define the template geometry, here simple 6~mm by 12~mm rectangles as schematically shown in Figure~\ref{fig:template_profiles}a and Figure~\ref{fig:templates_optical and depth}.  A development step then strips the photoresist from the regions to be etched and fully solidify the photoresist over the regions to be protected, before the etching step is finally performed. The depth of the template is controlled by the combination of etch rate, in turn a function of many factors including substrate material, etchant, concentrations, and temperature, and of etch duration. Last, the substrate is cleaned, including stripping of the photoresist, and inspected.

Templates were also fabricated by machining, enabling comparison to the etched templates. For this purpose, we machined templates as shallow pockets on the same 316L plate metal as the etched templates. Compared to etching, ordinary CNC machining is limited in its achievable depth precision, to the order of $\sim10\mu m$ \cite{kalpakjian2013manufacturing}. 
For reference, the manufacturing service provider Protolabs reports a standard machining tolerance of $\pm 130\mu m$, and for precision parts a standard precision tolerance of $\pm 50\mu m$ \cite{Protolabs2024}. With a typical nominal powder layer thickness of $\leq 100\mu m$, these tolerances make consistent template manufacturing by machining challenging.

The comparable templates in this work were machined on a HAAS VF-2SS, a 3-axis vertical machining center. Iteration of machining parameters was required to achieve the desired depth (measured with a depth micrometer). One additional challenge with machined templates is the required deburring of the very shallow pocket on a flat surface. The edges of the template were deburred by delicately filing the surface. Deburring is important because the burrs can interfere with the powder spreading as discussed later (e.g., a blade hitting the burr). Depth and surface profiles of the templates were measured with an optical profilometer (Keyence VK-X). 



\subsection{Single-layer LPBF experiments}
\label{subsec:Experimental setup}

Following template fabrication and measurement, single-layer LPBF experiments are performed by manually spreading powder into each template, and then melting desired areas of the template using a scanning laser in a commercial LPBF system. A SS316 powder commonly employed in LPBF (Carpenter Technology) is used, with the following size distribution: D10~=~22.6$\mu m$, D50~=~36.1$\mu m$, and D90~=~56.0$\mu m$.

Prior to spreading, a small amount of powder is placed directly next to the templates on the unetched surface of the substrate; the amount of powder is chosen to be comfortably more than is needed to fill the templates. Spreading is then performed by hand with a 1/8 inch thick machinist parallel (blade). The blade is rested in front of the powder pile, parallel to the short edge of the template and then moved across the template in one continuous motion. Throughout the motion, the blade is held in contact with the substrate surface. Burrs or uneven edges that might occur on machined templates make this smooth spreading motion significantly more difficult as the blade hits the burr and 'jumps', causing a chatter-like behavior which can inhibit the uniformity of the layer. The spreading motion on the etched templates is much smoother. 
Layer uniformity and packing density of the spread layer depend on the spreading velocity, tool geometry, and other factors. As described, a standard blade as used in most LPBF systems, and low (manual) speed is used to apply the powder, in an effort to resemble typical conditions. Alternatively, spreading could also be done with a suitable mechanized apparatus in a LPBF system.

\subsection{X-ray microscopy of powder layers}

\begin{figure}
 \begin{center}
   \includegraphics[width=0.45\textwidth]{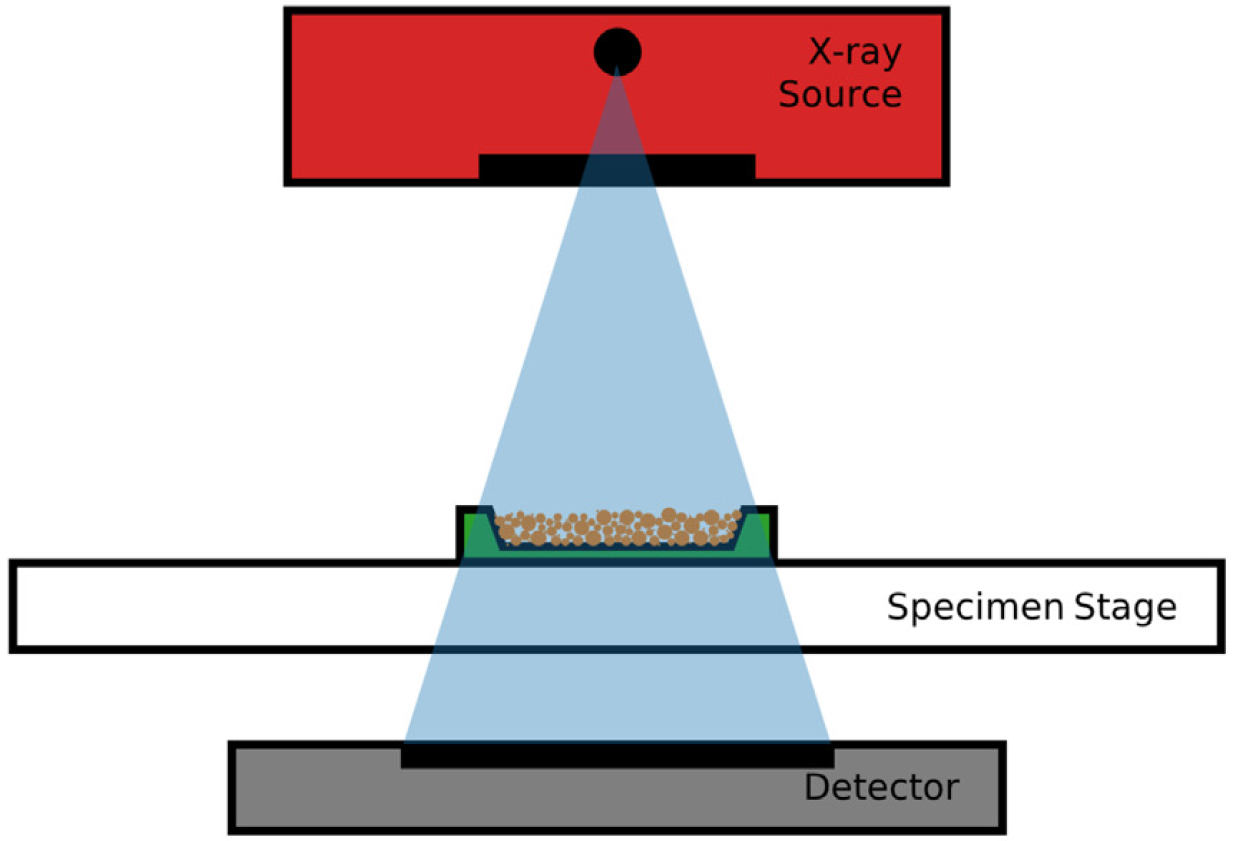}
 \end{center}
 \caption{Schematic visualization of the employed X-ray imaging setup for powder layers; Figure taken from~\cite{Penny2021}}
 \label{fig:x-ray schematic}
\end{figure}

Powder layers are imaged using a top-view X-ray transmission technique shown schematically in Figure~\ref{fig:x-ray schematic} and described in detail in~\cite{Penny2021} and previously also used in~\cite{PENNY2024Blade, PENNY2024Roller}. This technique enables us to spatially view the packing density and effective thickness of each layer, correlating template depth to layer quality and uniformity. The setup uses a Hamamatsu L12161-07 X-ray source, a Varex 1207NDT flat-panel detector, and a movable specimen stage where the substrate plate with the templates is placed. By varying the distances between X-ray source, specimen stage, and detector, the field of view and the resolution of the images can be set. The resolution is set to be approximately 18~$\mu m^2$ per pixel in a trade-off between a larger field of view to capture multiple wells in one image, and a fine image resolution. Each image taken is created by summing over 36 frames, after subtracting dark current frames taken after every 6 images to account for noise and create a high fidelity image. The exposure integration time is 90,000~ms, and the source is set to 50~keV and 200~$\mu$A.

An image of the substrate with the templates is taken without powder, before spreading. After spreading, images with powder are taken, and each image is analyzed by taking the ratio of the transmission data of the image with powder relative to the image without powder. Finally, these transmission measurements are translated into an effective depth field of the powder layer, which represents the thickness of the material if it were fully densified. The radiation transport model used to translate the radiation measurements into effective depth values is described in detail in~\cite{Penny2021}.

\subsection{Single-layer LPBF experiments}

\begin{figure}[ht!]
 \begin{center}
   \includegraphics[scale=1, keepaspectratio=true, width=.9\textwidth]{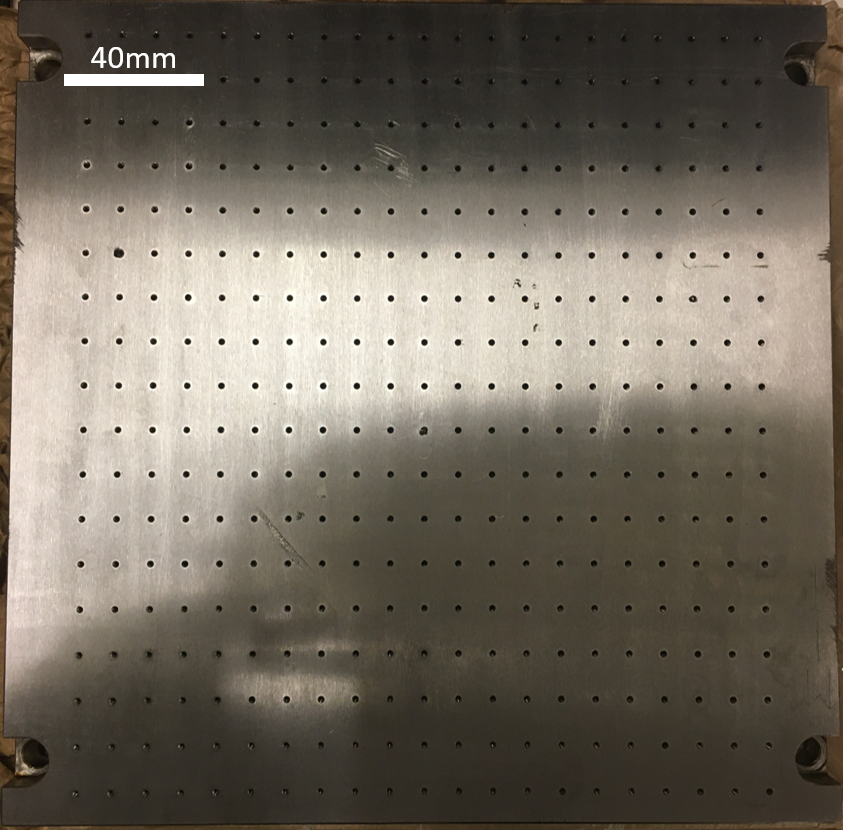}
 \end{center}
 \caption{EOS M290 build plate with threaded hole pattern to accommodate a large number of single layer templates. The hole pattern is optimized for flexibility and can hold up to 21 specimens with six rectangular templates each, allowing efficient large-scale parameter studies.}
 \label{fig:build plate}
\end{figure}

Single-layer, track-wise melting experiments are performed using a commercial LPBF machine (EOS M290), with a nominal calibrated laser spot size of $4\sigma=100~\mu m$. The templates used for the melting experiments have a nominal depth of 60~$\mu m$, 90~$\mu m$, and 120~$\mu m$, and the powder layer is prepared as described above. For all three depths, a full factorial study is performed, with scan speeds of $v~=~$\{500, 1000\}~mm/s, laser power levels of P~=~\{75, 100, 125, 150, 175, 200, 225, 250, 275, 300, 325, 350\}~W.  Each parameter combination is used to create 5 lines of 6~mm length with 1~mm center-center spacing. Each template accommodated two parameter combinations. Thus, the 24 parameter combinations described above required 12 templates. A standard build plate was prepared with a threaded hole pattern to allow locating and securing a total of 21 substrate strips on the build plate at the same time (see Figure~\ref{fig:build plate}). This setup allows for efficient processing of large arrays of parameters.


\subsection{Melt track characterization}

\begin{figure}[h!]
 \begin{center}
   \includegraphics[scale=1, keepaspectratio=true, width=\textwidth]{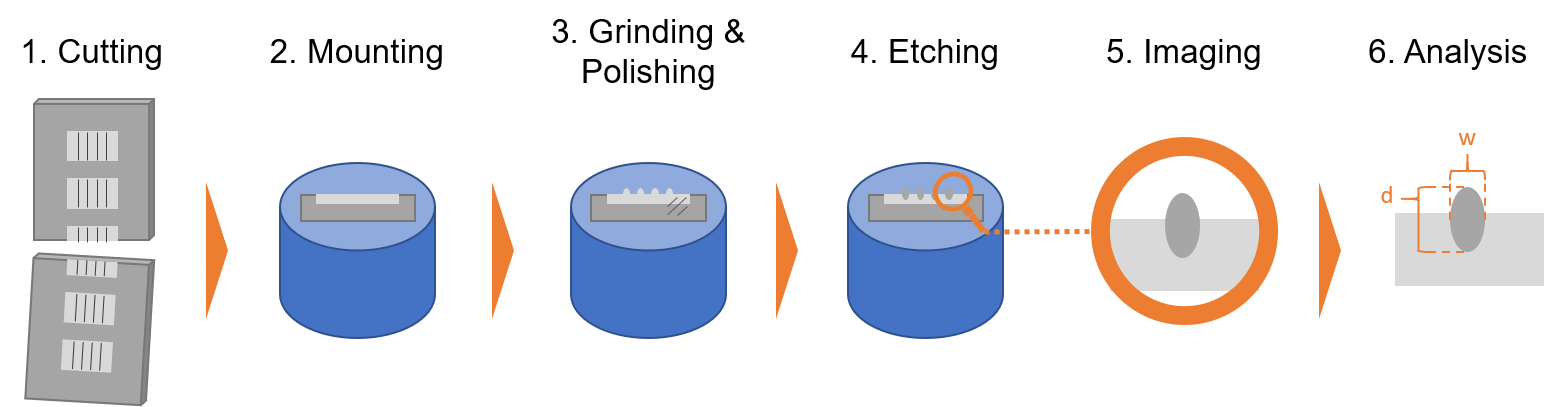}
 \end{center}
 \caption{Schematic of the melt track characterization process}
 \label{fig:cs analysis process}
\end{figure}

After laser melting, each strip is post-processed in order to analyze the melt track cross-section, as described schematically in Figure~\ref{fig:cs analysis process}. First, the strips are cut perpendicular to the melt track direction. The width of the strip is designed to be barely wider than the width of the templates in order to minimize cutting time. Each template (and thus melt track) is cut twice in order to increase the amount of data points extracted per experiment. The cut is performed with a Buehler Isomet 4000 precision saw with a diamond blade operated at 4000~rpm and a cutting speed of 6~mm/min. Subsequently, the cut is deburred and the slice is mounted in a cylindrical puck with a diameter of 1.25~in with a Buehler Simplimet 4000 compression hot mounting machine. Two slices are mounted in each puck.

\begin{figure}[t]
 \begin{center}
   \includegraphics[width=0.47\textwidth]{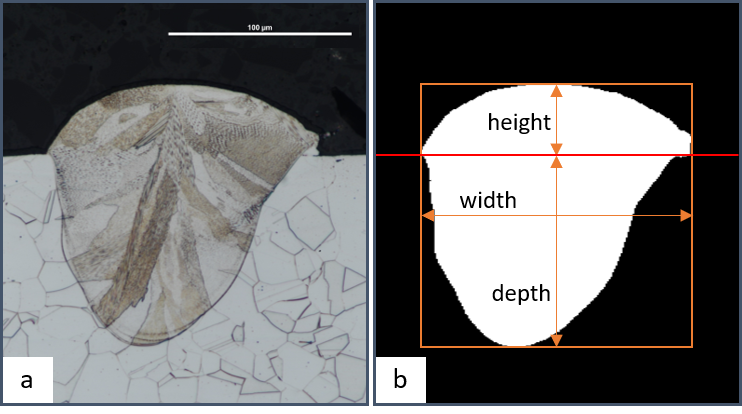}
 \end{center}
 \caption{Exemplary cross-section of melt track generated with a layer thickness of 60$\mu m$, laser power of 325~W, and a scan speed of 1000~mm/s: (a) optical image, (b) image mask generated by image analysis tool (see \cite{Shah2024_image, Shah2024_github})}
 \label{fig:cs_mask_example}
\end{figure}

Next, the samples are ground with increasingly finer grit size on a Struers Tegrapol automated grinding machine, first with a grit 320 for 2 minutes and then with a grit 800 for two times 2~minutes. A sample holder that can fit six pucks at the same time is used on the grinding head, allowing to process all samples from one substrate strip at the same time. The force normal to the grinding surface is 25~N per sample and the grinding plate speed is set to 300~rpm with the head co-rotating with 60~rpm. 

After grinding, two polishing steps are performed on a Buehler AutoMet 250 Pro automated polishing machine in order to get a mirror-finish on the samples. First, a 3~$\mu m$ diamond suspension is used, and the final polishing step is done with a 0.05~$\mu m$ alumina suspension. Both polishing steps are run for 3~minutes with the polishing head counter-rotating to the polishing pad. The normal force is 27~N per sample, the polishing pad rotates at 300~rpm and the polishing head is at 60~rpm. The samples are cleaned in an ultrasonic cleaner between each grinding and polishing step.

After a mirror-finish is achieved, the samples are etched with diluted Aqua Regia (equal parts water, nitric acid, and hydrochloric acid). The etching is done by submerging the surface in the etchant for 30-45~s, subsequently briefly submerging/rinsing in deionized water and finally washing with ethanol to avoid water stains. It is critical that the etching is done immediately after the final polishing step because the SS316 is prone to instantaneously forming a passivation (oxide) layer on the polished surface which impedes the etching process.

After etching, the melt track cross-sections are imaged using a Nikon Eclipse MA200 inverted microscope. The images are then analyzed using an automated tool that identifies the shape of the melt track cross-section, creates a mask, and calculates the height and width (Figure~\ref{fig:cs_mask_example})~\cite{Shah2024_image, Shah2024_github}. The computer vision algorithm is based on a U-net architecture. It utilizes a convolutional neural network (CNN) that first reduces a cross-sectional image to a latent space representation (encoding) and then decodes that representation back to a 2D image. Each upsampling step on the decoding side is informed by information from the downsampling steps, which allows contextual information to be preserved.

The model was trained on a number of curated, highly diverse melt track cross-sections (real images as well as images created by image augmentation) in order to achieve a high level of accuracy across a wide range of shapes, colors, and levels of etching. The algorithm also detects the scale bar and uses optical character recognition (OCR) to read the real length of the scale bar. This reveals the conversion ratio or pixels to $\mu m$, which is applied to the measured width and depth of the melt pool.

\section{Results and discussion}

\subsection{Dimensional characteristics of etched and machined templates}

\begin{figure*}[ht!]
 \begin{center}
   \includegraphics[scale=1, keepaspectratio=true, width=\textwidth]{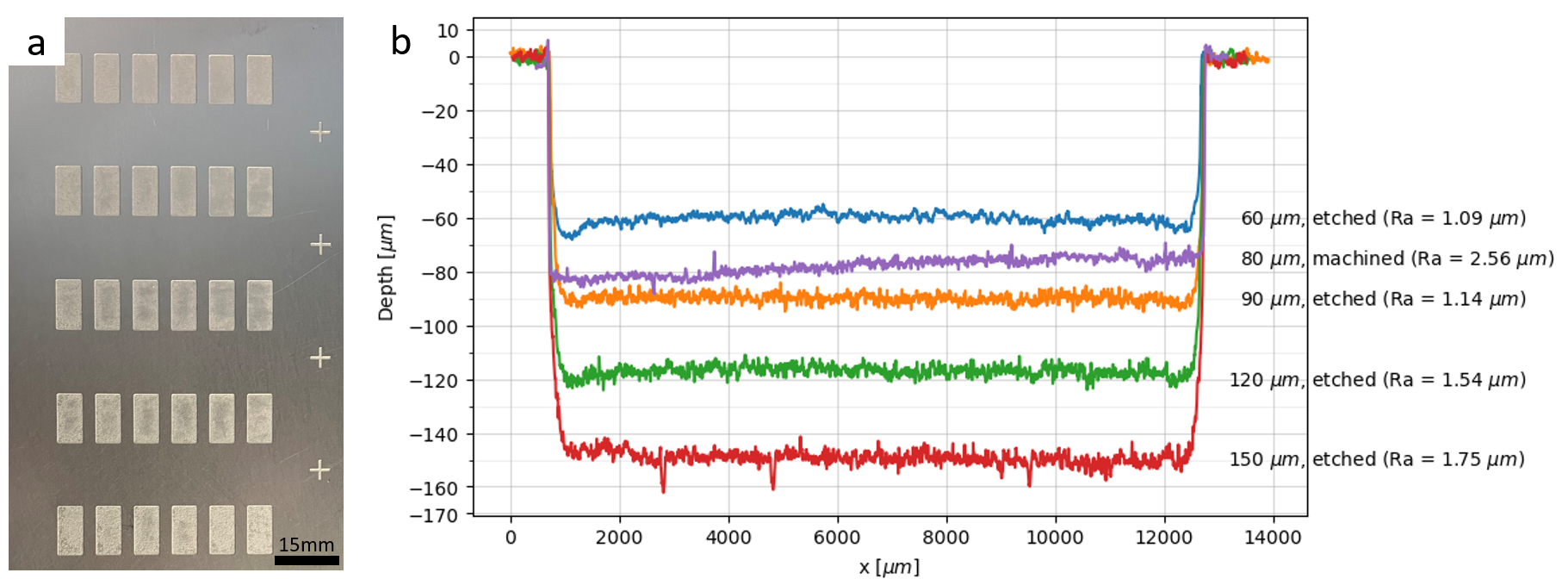}
 \end{center}
 \caption{a) Photo of exemplary templates etched into a SS316 sheet, and b) depth profile etched and machined templates in SS316}
 \label{fig:template_profiles}
\end{figure*}

\begin{figure*}[ht!]
 \begin{center}
   \includegraphics[scale=1, keepaspectratio=true, width=\textwidth]{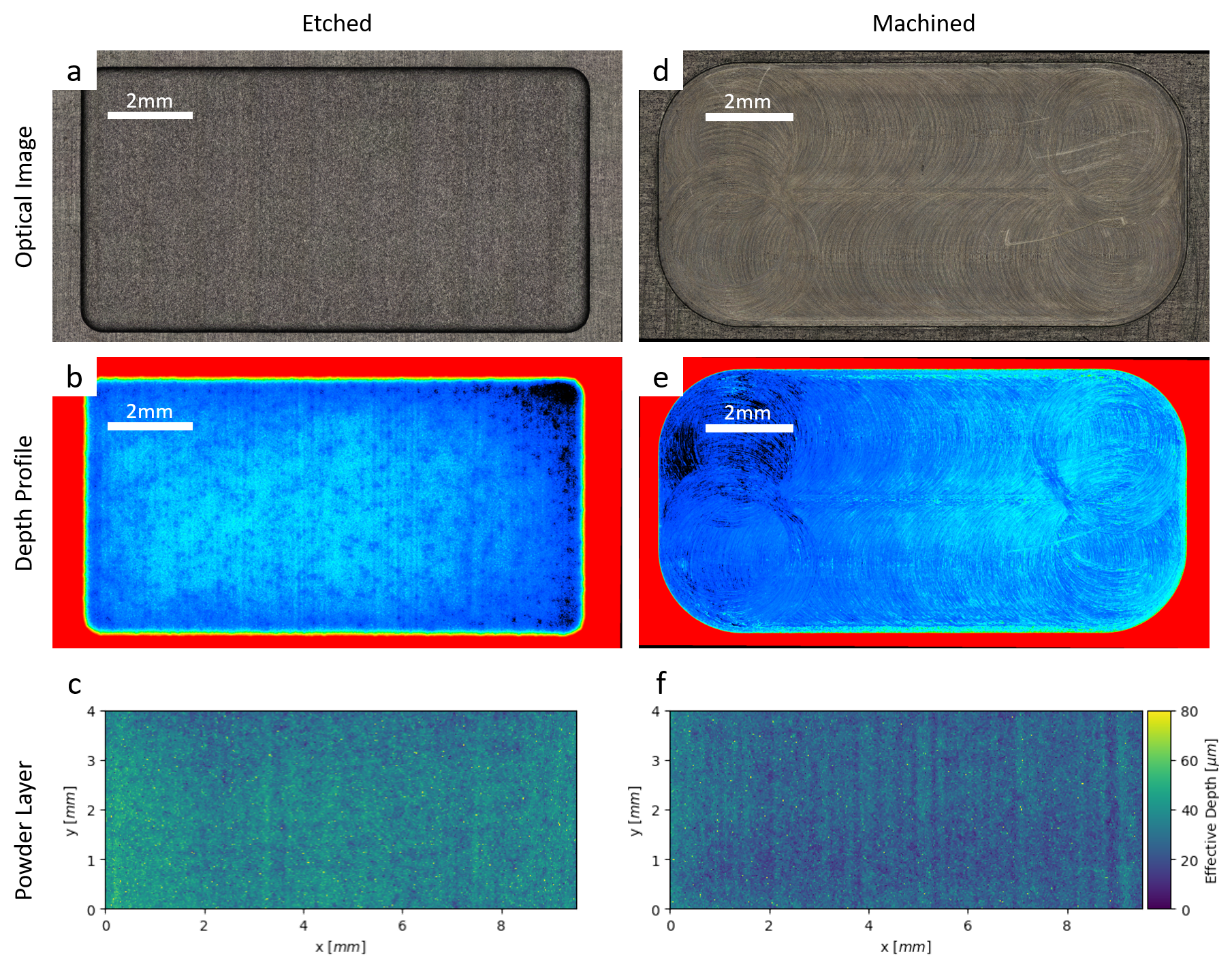}
 \end{center}
 \caption{Top-view images of exemplary templates: (a) optical image, (b) depth profile, and (c) X-ray transmission image of exemplary powder layer in etched template (90~$\mu m$ depth); (d) optical image, (e) depth profile, and (f) X-ray transmission image of exemplary powder layer in machined template (80~$\mu m$ depth) with pronounced spreading chatter, likely a result of the blade sliding over residual burrs; Notes: Spreading direction from left to right; Color-scale of depth profile (b and e) set to display range of approximately 37~$\mu m$ for improved visualization of surface texture}
 \label{fig:templates_optical and depth}
\end{figure*}

Figure~\ref{fig:template_profiles}b shows the depth profiles of exemplary etched ($60\mu m$, $90\mu m$, $120\mu m$, and $150\mu m$ nominal depth) and machined ($90\mu m$ nominal depth) templates. The etched templates have a high accuracy of holding the nominal depth across the entire area of the template, with slightly deeper areas near the edges. The machined template with a nominal depth of $90\mu m$ has a depth of approximately $80\mu m$, and further is less consistent across the length of the template in that depth. The machined template that was measured in Figure~\ref{fig:template_profiles}b is the same template displayed in Figure~\ref{fig:templates_optical and depth}. The machining error of $10\mu m$ is comparatively low in the context of achievable machining precision as described above. For clarity, the machined template with a nominal depth of $90\mu m$ will be referred to as $80\mu m$ depth going forward.

Roughness was measured along lines (10~mm length) traversing the templates, to exclude edge effects. Etching creates a lower surface roughness than machining, and the surface roughness increases with depth for the etched templates: Ra of $1.09\mu m$, $1.14\mu m$, $1.54\mu m$, and $1.75\mu m$ for etched $60\mu m$, $90\mu m$, $120\mu m$, and $150\mu m$ template respectively, as well as $2.56\mu m$ for the machined template (see Figure~\ref{fig:template_profiles}b). The structure of the measured surface roughness also differs between the etched and the machined templates. In Figure~\ref{fig:templates_optical and depth}, both the optical image as well as the depth map show that the surface texture is unstructured for the etched template, while the tool path is visible on the machined template. The machined template has a small scratch on the right side, that likely stems from the deburring operation. The color-scale of the height map is set to display a range of approximately $37\mu m$ for both images in order to improve the visualization of the surface texture.

\subsection{Powder layer quality in single-layer templates}

X-ray microscopy is used to investigate the quality of powder layers spread into the templates. Here, etched and machined templates with depths of $90\mu m$ and $80\mu m$ respectively are compared. The X-ray images were cropped by 1~mm from each side to exclude edge effects, leaving an analyzed surface area of approximately 10~mm $\times$ 4~mm per template. 

Exemplary images of powder layers spread into the templates are shown in Figure~\ref{fig:templates_optical and depth}c and \ref{fig:templates_optical and depth}f, where the spreading direction was from left to right. For the etched template, the layer is relatively uniform, while exhibiting slight variation in effective depth along with some streaks parallel to the spreading direction.  In the machined template, the effect of pronounced chatter during spreading is visible via vertical lines normal to the spreading direction. The image is purposefully chosen to show the effect of blade chatter on the powder layer -- not all layers exhibit this effect to the same extent. Stripes due to chatter were also visible in some of the layers spread into the etched template. This is an artifact of manual spreading with a blade and is likely caused by burrs on the machined template and particles getting caught between the template edge and the blade. 

The average effective depth (as defined in~\cite{Penny2021}) of the deposited powder layers is 34.9~$\mu m$ for the etched template and 27.1~$\mu m$ for the machined template. These values represent averages of the depth of 15 layers spread and measured from X-ray images of each template. Overall, these values are consistent with the findings of Penny and Hart in \cite{Penny2021}, who evaluated the relationship between effective depth and template depth, for etched silicon templates. The average layer depth in the etched template represents a volumetric packing density of 39\%. The lower packing density in the machined template (34\%) can be attributed to its surface roughness which impedes layer uniformity, the lower relative template depth, and the aforementioned propensity for chatter during manual spreading.

\subsection{Single-track LPBF experiments in etched templates}

To demonstrate the utility of the metal templates for exploring the scaling of LPBF parameters, single-track parameter studies were performed as outlined in Section~\ref{subsec:Experimental setup}. Three depths of etched templates (60~$\mu m$, 90~$\mu m$, 120~$\mu m$) were chosen as these span a range of equivalent layer thicknesses (considering effective powder layer depth) that are commonly used in LPBF practice.  As the template depth sets the powder layer thickness, the effective layer thickness that is created by the consolidation of the powder in LPBF melting~\cite{mindt2016powder} will always be significantly less than the template depth. For example a 100~$\mu m$ template depth single layer melting experiment resembles a powder layer thickness that would be found in a LPBF build with 40~$\mu m$ nominal layer thickness and 40\% packing fraction.

\begin{figure*}
 \begin{center}
   \includegraphics[scale=1, keepaspectratio=true, width=\textwidth]{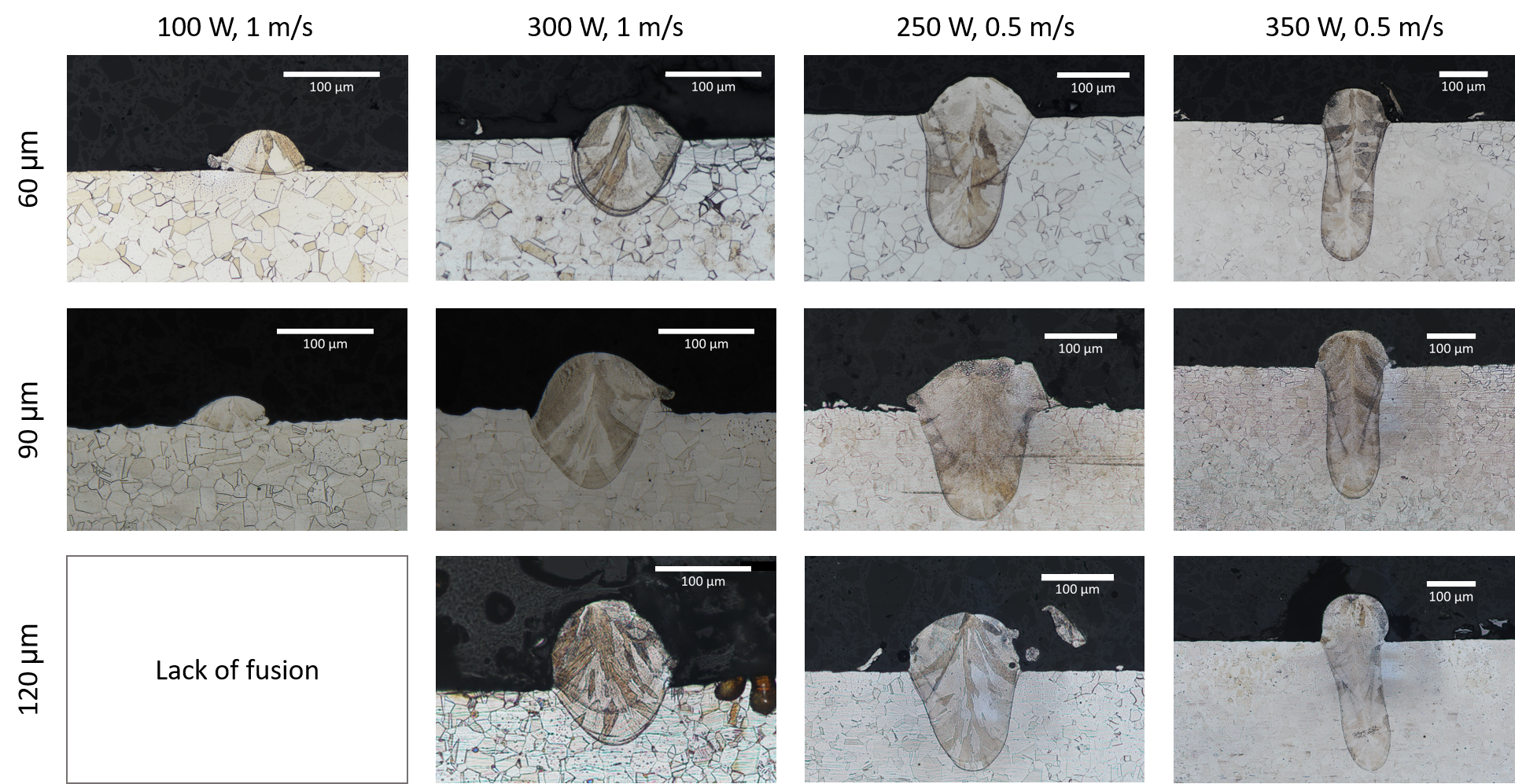}
 \end{center}
 \caption{Exemplary cross-sectional images of single-track experiments}
 \label{fig:cs examples}
\end{figure*}

Cross-sectional imaging of the templates with melt tracks shows a characteristic beaded morphology with, under typical conditions, increasing penetration into the substrate (beneath the bottom of the etched surface) with increasing energy density, as shown in Figure~\ref{fig:cs examples}. These single tracks are a convenient means to understand scaling as related to fundamental process physics. Scaling of track dimensions with the parameter normalized enthalpy has been applied in several recent studies. Starting from Hann et al.~\cite{hann2011simple} and building on King et al.~\cite{king2014observation}, Rubenchik et al.~\cite{rubenchik2018scaling} and Naderi et al.~\cite{naderi2023fidelity} observed a linear scaling of the normalized melt track depth with normalized enthalpy with a slope of approximately 0.25. 

Normalized enthalpy in this context is defined as:
\begin{align}
    \frac{\Delta H}{h_s} &= \frac{A P}{\rho c_{heat} (T_{melt} - T_0) \sqrt{\pi \alpha v \sigma^3} },
\label{eq:norm_enthalpy}
\end{align}
with laser absorptivity $A=0.4$, laser power $P$ as defined in Section~\ref{subsec:Experimental setup}, density $\rho=8000 \frac{kg}{m^3}$, specific heat $c_{heat} = 500 \frac{J}{kg~K}$, melt temperature $T_{melt} = 1673 K$, substrate temperature $T_0 = 293 K$, thermal diffusivity $\alpha = 4.675\times 10^{-6} \frac{m^2}{s}$, laser scan speed $v$ as defined in Section~\ref{subsec:Experimental setup}, and the variance of the Gaussian laser intensity profile $\sigma=25\times 10^{-6} m$ (i.e., the commonly reported laser spot size diameter is $4\sigma$).
The normalized depth is defined as:
\begin{align}
    \frac{d}{\delta} &= \frac{d}{\sqrt{\frac{\alpha \sqrt{2}\sigma}{v} }},
\label{eq:norm_depth}
\end{align}
with the measured depth of the melt pool $d$, corresponding to depth (below the substrate surface) plus height (above the substrate surface) as identified in Figure~\ref{fig:cs_mask_example}.

\begin{figure}
     \begin{subfigure}{0.45\textwidth}
         \centering
         \includegraphics[trim = {0mm 0mm 8mm 10mm}, clip, scale=1, keepaspectratio=true, width=\textwidth]{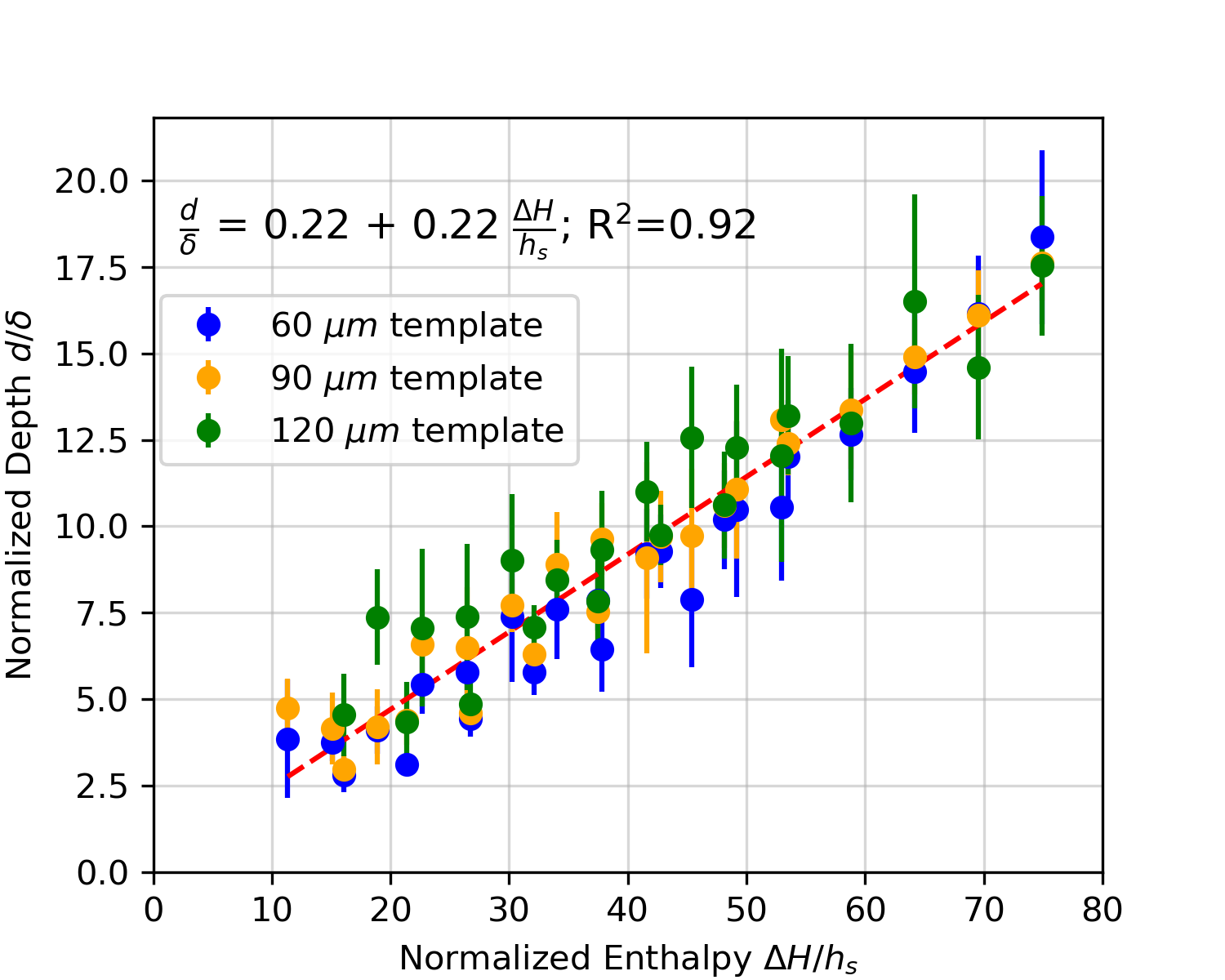}
         \caption{}
         \label{fig:norm_enthalpy plot}
     \end{subfigure}
     \vfill
     \begin{subfigure}{0.45\textwidth}
         \centering
         \includegraphics[trim = {0mm 0mm 8mm 10mm}, clip, scale=1, keepaspectratio=true, width=\textwidth]{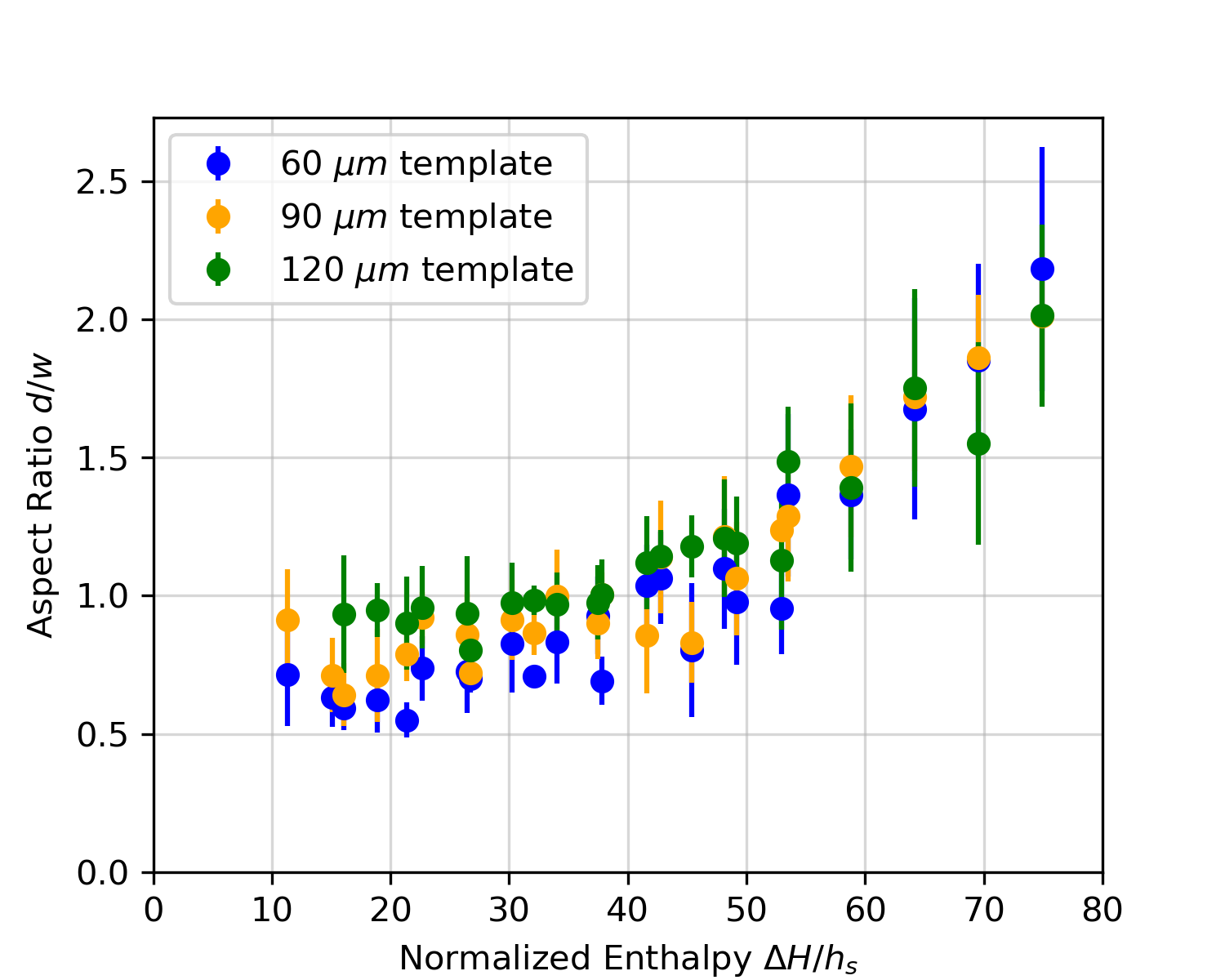}
         \caption{}
         \label{fig:aspect ratio}
     \end{subfigure}
     \vfill
     \begin{subfigure}{0.45\textwidth}
         \centering
         \includegraphics[trim = {0mm 0mm 8mm 10mm}, clip, scale=1, keepaspectratio=true, width=\textwidth]{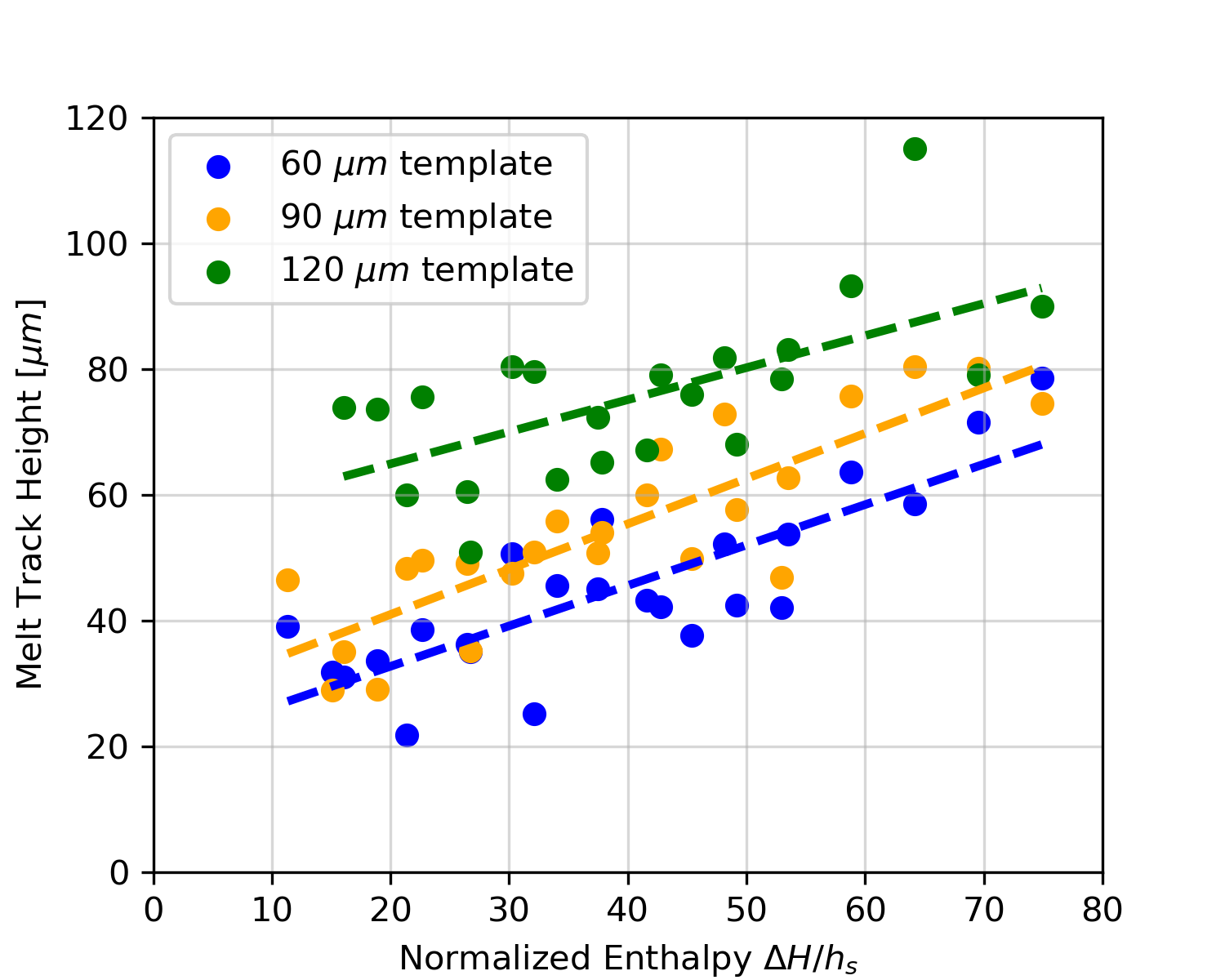}
         \caption{}
         \label{fig:track height plot}
     \end{subfigure}
     \caption{Melt track dimensions, measured by cross-section image analysis, and plotted against normalized enthalpy: a) normalized depth, b) aspect ratio, and c) melt track height. Each point reflects the average of the cross-sections for a given parameter combination and template depth, with the errorbar indicating the standard deviation of that sample of data points (average of ~8 cross-sections analyzed per parameter combination and depth).}
     \label{fig:cross-section results}
\end{figure}

In our experiments with single-layer templates, the chosen laser parameters cover a range of 10~$<\frac{\Delta H}{h_s}<$~80, with the objective of observing all common laser melting modes (lack-of-fusion, conduction mode, keyhole mode).
Figure~\ref{fig:norm_enthalpy plot} shows the experimental data plotted as normalized depth vs.~normalized enthalpy, including a linear fit in red that follows the relation $\frac{{d}}{\delta} = 0.22 + 0.22\frac{\Delta H}{h_s}$, with a fit of R$^2$=0.92. The slope of 0.22 agrees with the findings of  \cite{naderi2023fidelity, rubenchik2018scaling}, and despite local fluctuations in the data, for each template depth the entire range is captured well by a single linear fit.

As described earlier, the study is fully factorial across 3 template depths, 2 scan speeds, and 12 laser power levels, with an average of 8 cross-sectional images analyzed per parameter combination. The results are then averaged per parameter combination and template depth and the standard deviations are indicated with bars. The uniformity of scaling across this wide range of parameters and different layer thicknesses is particularly notable, as it also shows that the scaling law is robust to layer thickness -- a parameter not considered in the derivation. 

Exemplary melt track cross-sections are depicted in Figure~\ref{fig:cs examples}. On the low end of energy density, the images for P=100~W, $v=1$~m/s correspond to a normalized enthalpy value of 15. The images are examples of a cross-sections where almost no substrate material is melted, likely a case that would result in lack-of-fusion failure a in three-dimensional build for the layer thicknesses 60~$\mu m$ and 90~$\mu m$ -- for the layer thickness of 120~$\mu m$ no material is deposited on the substrate. The parameter combination P=300~W, $v=1$~m/s corresponds to a normalized enthalpy value of 45 and is an example of an almost spherical melt track cross-section, which is characteristic of conduction mode. The transition to keyhole mode melt tracks is visible for the parameter combination P=250~W, $v=0.5$~m/s, corresponding to a normalized enthalpy value of 54, where the bottom of the melt pool has near-vertical wall angles. Based on examination of the images, the transition from circular conduction morphology to keyhole morphology occurs around a normalized depth value of 10 and a normalized enthalpy of around 50. Pronounced keyhole morphology is visible for the parameter combination P=350~W, $v=0.5$~m/s, where the melt track is very deep for a normalized enthalpy value of approximately 75.

More generally, the melt track geometry of a single-track experiment is relatively independent of the powder layer thickness, given that the relationship between normalized depth and enthalpy for all three layer thicknesses can be fit by a single line in Figure~\ref{fig:norm_enthalpy plot}, as well the similarity of the progression of aspect ratio in Figure~\ref{fig:aspect ratio}. Previous studies have shown that melt track depth scales similarly with and without powder (bare plate) when considering the effective thickness of the powder layer~\cite{bogdanova2024mesoscale,cunningham2019keyhole}. Figure~\ref{fig:track height plot} shows the melt track height above the substrate -- increasing steadily with increasing normalized enthalpy with a height offset with increasing layer thickness. This means that adding more powder without changing laser parameters successively moves the processing into the lack-of-fusion (LOF) regime as the deposited material increases and the remelting depth decreases. In the data, this happens for the two lowest normalized enthalpy levels and the 120~$\mu m$ template (missing data points of the green curve in Figure~\ref{fig:norm_enthalpy plot}). As a result, with an increasing layer thickness and while keeping the spot size constant, the LOF-barrier moves to higher normalized enthalpy values. The keyhole transition criteria, however, stays unchanged around a normalized enthalpy value of 50 (see analysis of aspect ratios in Figure~\ref{fig:aspect ratio}). This means that the window to operate between LOF and keyholing is narrowing with increasing layer thickness.


\begin{figure*}
 \begin{center}
   \includegraphics[scale=1, keepaspectratio=true, width=\textwidth]{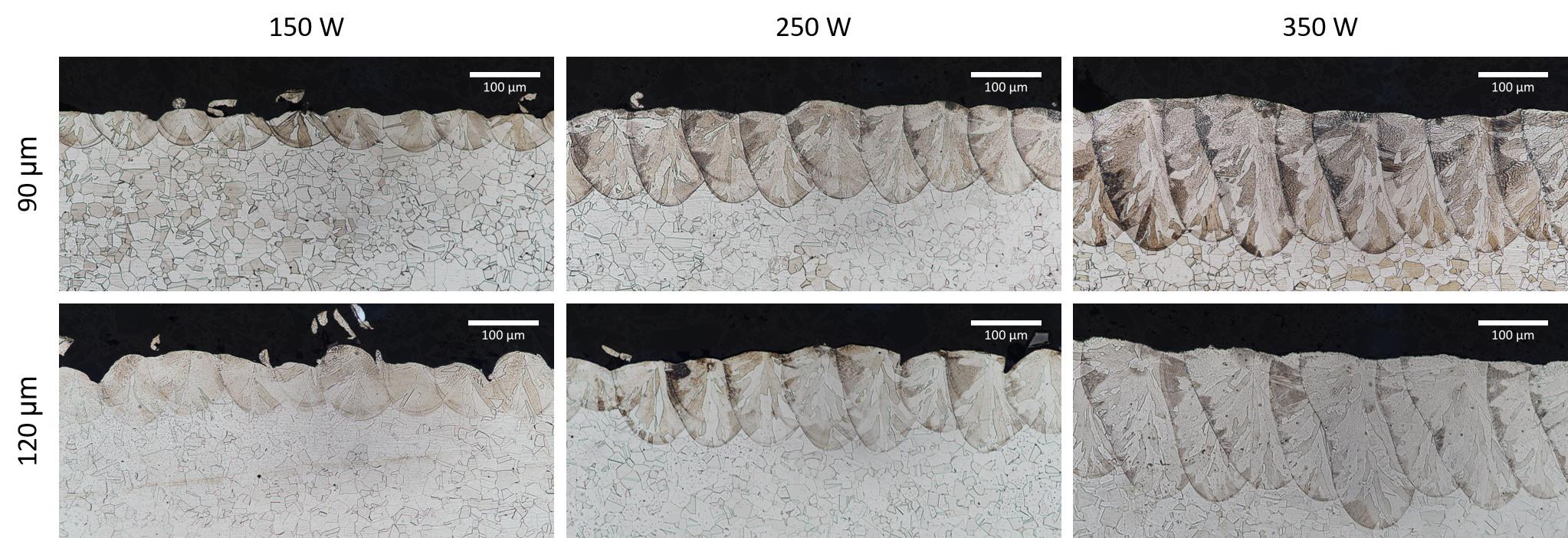}
 \end{center}
 \caption{Exemplary cross-sectional images of raster scanning experiments with scan speed of $v=1$~m/s and hatch spacing of 80~$\mu m$}
 \label{fig:raster examples}
\end{figure*}

Figure~\ref{fig:raster examples} shows exemplary raster scanning experiments using template depths of 90~$\mu m$ as well as 120~$\mu m$, and the same laser parameters as outlined in Section~\ref{subsec:Experimental setup}. Random sampling of melt track depths in the raster scans confirmed the melt track depths measured for single line melting experiments. For laser power P$<150$~W, the raster scan surface is not continuous due to the melt track width being smaller or similar to the hatch spacing, or LOF defects. This threshold corresponds to a normalized enthalpy of 23. Smoothness of the surface and consistency of the melt track cross-sections increases continuously with increasing laser power. For the highest applied power of 350~W, the transition from conduction mode to keyhole mode is visible, but no keyhole pores were detected in the cross-sections with a normalized enthalpy value of 53. 

The raster scanning results are consistent with the single-track results, indicating that single-track experiments can be used as a good proxy for raster scanning results, and likely for full LPBF builds. In such an experimental design, the hatch spacing for a full LPBF build can be determined based on the measured melt track width of the single-track results, and the layer thickness in combination with the observed melt track depth.
More generally, the precision template approach proves to be convenient and accurate for studying the scaling of LPBF parameters in corroboration with previous studies, and could be applied to a wide variety of materials beyond steel. For example, studies of Ti alloys~\cite{brudler2024systematic}, Al alloys~\cite{bogdanova2024mesoscale, ghasri2023single}, and Nickel-based alloys~\cite{simson2024experimental} commonly use etched cross-sectional images to understand the melting behavior.
The methodology could be adapted to mechanized spreading, e.g. by referencing the top surface of the template build plate to a (compliant) blade or roller, after carefully leveling the build plate. 
In addition, generating a large number of cross-sections can facilitate a more informed understanding of how melt track geometry changes with key process parameters. This may be of particular interest for exploring new combinations of powder characteristics and layer dimensions (e.g., very thin or thick layers, with commensurate changes in the powder size distribution), intricate geometries with tailored laser spot sizes~\cite{kosiba2025fabrication}, and for beam shaping approaches that appear to improve melt pool stability and throughput~\cite{yuan2022understanding, grunewald2022flexible, holla2023laser}.

\section{Conclusion}

The demonstrated workflow for single-layer LPBF experimentation using precision metal template substrates can facilitate versatile experimentation with new materials, as well as development and verification of process parameter scaling behavior. Wet etching which is applicable to many metal alloys, enables precision control of template depth; however, precision machining of templates can be applied to an even wider variety of materials while sacrificing surface quality and depth accuracy. In this study, use of etched SS316 templates to accommodate controlled thickness powder layers allowed us to explore variation in melt track geometry and dimensions over a wide range of representative process conditions, and elucidate how transitions between process modes change with the normalized enthalpy value dictated by the laser scan parameters. The availability of a large number of melt track cross-section images prepared under well-controlled conditions can also facilitate new understanding of process boundaries, and assist calibration of computational models in future work.
\section*{Acknowledgements}
\label{sec:Acknowledgements}

Financial support was provided by a MathWorks MIT Mechanical Engineering Fellowship (to R.W.), the National Science Foundation (Award EEC-1720701, subcontracted to MIT by the University of Illinois at Urbana-Champaign), and Honeywell Federal Manufacturing \& Technologies (FM\&T). We thank Christoph Meier for insightful discussions and support for this research. Further, we thank Bethany Lettiere and United Western Enterprises for guidance on etching of templates. We also thank Shaymus Hudson from the Laboratory for Physical Metallurgy at MIT, as well as Dave Follette from the Advanced Digital Design and Fabrication (ADDFab) facility at the University of Massachusetts Amherst for their support with use of shared facilities and related techniques.  Last, we appreciate collaboration with Aagam Shah, Elif Ertekin, and Sameh Tawfick of the University of Illinois at Urbana-Champaign, with whom we co-developed the image analysis tool that is utilized for this study and published separately.

\section*{Conflict of Interest Statement}
On behalf of all authors, the corresponding author states that there is no conflict of interest.

\onecolumn

\newpage
\bibliographystyle{elsarticle-num} 
\bibliography{refs.bib}

\newpage
\beginsupplement
\section{Schematic of melt track characterization process}


\newpage
\section{Build plate with hole pattern for large scale parameter study}

\end{document}